\documentclass[12pt,preprint]{aastex}
\usepackage{appendix}
\usepackage{lscape}

\slugcomment{second revised draft, submitted 13 April 2010}

\begin{document}{}

\def\t0{\theta_{\circ}}
\def\muo{\mu_{\circ}}
\def\sd{\partial}
\def\be{\begin{equation}}
\def\en{\end{equation}}
\def\bv{\bf v}
\def\bvo{\bf v_{\circ}}
\def\ro{r_{\circ}}
\def\rhoo{\rho_{\circ}}
\def\etal{et al.\ }
\def\msun{\,M_{\sun}}
\def\rsun{\,R_{\sun}}
\def\rstar{\,R_{star}}
\def\lsun{L_{\sun}}
\def\mbol{M_{bol}}
\def\msunyr{{\it M_{\odot}} yr^{-1}}
\def\kms{\rm \, km \, s^{-1}}
\def\mdot{\dot{M}}
\def\mdotd{\dot{M}_{\rm d}}
\def\Md{\dot{M}}
\def\curf{{\cal F}}
\def\ecs{erg cm^{-2} s^{-1}}
\def \haebe{HAeBe}
\def \mum {\,{\rm \mu m}}
\def \simali {{\sim\,}}
\def \K {\,{\rm K}}
\def \Angstrom     {\,{\rm \AA}}
\newcommand \g            {\,{\rm g}}
\newcommand{\ltapp}{\raisebox{-.4ex}{\rlap{$\sim$}} \raisebox{.4 ex}{$<$}}
\newcommand{\gtapp}{\raisebox{-.4ex}{\rlap{$\sim$}} \raisebox{.4 ex}{$>$}}
\newcommand \cm           {\,{\rm cm}}

\title{The Mass-Loss Return from Evolved Stars to the Large Magellanic 
Cloud II: Dust Properties for Oxygen-Rich Asymptotic Giant Branch Stars}

\author{Benjamin A. Sargent\altaffilmark{1},
S. Srinivasan\altaffilmark{2},
M. Meixner\altaffilmark{1},
F. Kemper\altaffilmark{3},
A.~G.~G.~M. Tielens\altaffilmark{4},
A.~K. Speck\altaffilmark{5},
M. Matsuura\altaffilmark{6,7},
J.-Ph. Bernard\altaffilmark{8},
S. Hony\altaffilmark{9},
Karl D. Gordon\altaffilmark{1},
R. Indebetouw\altaffilmark{10,11},
M. Marengo\altaffilmark{12},
G.~C. Sloan\altaffilmark{13},
Paul M. Woods\altaffilmark{3}
}

\altaffiltext{1}{Space Telescope Science Institute, 3700 San Martin 
                 Drive, Baltimore, MD 21218, USA;
                 {\sf sargent@stsci.edu}}
\altaffiltext{2}{Institut d'Astrophysique de Paris, 98 bis, Boulevard Arago, 
                 Paris 75014, France}
\altaffiltext{3}{Jodrell Bank Centre for Astrophysics, Alan Turing Building, 
                 School of Physics and Astronomy, The University of Manchester, 
                 Oxford Road, Manchester, M13 9PL, UK}
\altaffiltext{4}{Leiden Observatory, P.~O. Box 9513, NL-2300 RA Leiden, 
                 The Netherlands}
\altaffiltext{5}{Physics \& Astronomy Department, University of Missouri, 
                 Columbia, MO 65211}
\altaffiltext{6}{Institute of Origins, Department of Physics and Astronomy, 
                 University College London, Gower Street, London WC1E 6BT, UK}
\altaffiltext{7}{Institute of Origins, Mullard Space Science Laboratory, 
                 University College London, Holmbury St. Mary, Dorking, Surrey 
                 RH5 6NT, UK}
\altaffiltext{8}{Centre d'\'{E}tude Spatiale des Rayonnements, 9 Av. du Colonel 
                 Roche, BP 44346, 31028 Toulouse cedex 4, France}
\altaffiltext{9}{Laboratoire AIM, CEA/DSM - CNRS - Universit\'{e} Paris Diderot 
                 DAPNIA/Service d'Astrophysique B\^{a}t. 709, CEA-Saclay F-91191 
                 Gif-sur-Yvette C\'{e}dex, France}
\altaffiltext{10}{Department of Astronomy, University of Virginia, P.O. Box 
                 400325, Charlottesville, VA 22904}
\altaffiltext{11}{National Radio Astronomy Observatory, 520 Edgemont Road, 
                 Charlottesville, VA 22903}
\altaffiltext{12}{Department of Physics and Astronomy, Iowa State University, 
                 Ames, IA 50011, USA}
\altaffiltext{13}{Department of Astronomy, Cornell University, Ithaca, NY 
                 14853}

\begin{abstract}

We model multi-wavelength broadband {\it UBVIJHK$_{\rm s}$} and {\it Spitzer} 
IRAC and MIPS photometry and IRS spectra from the SAGE and SAGE-Spec 
observing programs of two oxygen-rich asymptotic giant branch (O-rich AGB) 
stars in the Large Magellanic Cloud (LMC) using radiative transfer models of 
dust shells around stars.  We chose a star from each of the bright and faint 
O-rich AGB populations found by earlier studies of the SAGE sample in order 
to derive a baseline set of dust properties to be used in the construction 
of an extensive grid of radiative transfer models of the O-rich AGB stars found 
in the SAGE surveys.  From the bright O-rich AGB population we chose HV 
5715, and from the faint O-rich AGB population we chose SSTISAGE1C 
J052206.92-715017.6 (SSTSAGE052206).  We found the complex indices of 
refraction of oxygen-deficient silicates from \citet{oss92} and a ``KMH''-like grain 
size distribution with $\gamma$ of -3.5, $a_{\rm min}$ of 0.01$\mum$, and 
$a_{\rm 0}$ of 0.1$\mum$ to be reasonable dust properties for these models.  
There is a slight indication that the dust around the faint O-rich AGB may be 
more silica-rich than that around the bright O-rich AGB.  Simple models of gas 
emission suggest a relatively extended gas envelope for the faint O-rich AGB 
star modeled, consistent with the relatively large dust shell inner radius for the 
same model.  Our models of the data require the luminosity of SSTSAGE052206 
and HV 5715 to be $\simali$5\,100$\lsun$ and $\simali$36\,000$\lsun$, 
respectively.  This, combined with the stellar effective temperatures of 3\,700\,K 
and 3\,500\,K, respectively, that we find best fit the optical and near-infrared data, 
suggests stellar masses of $\simali$ 3$\msun$ and $\simali$7$\msun$.  
This, in turn, suggests that HV 5715 is undergoing hot bottom burning and that 
SSTSAGE052206 is not.  Our models of SSTSAGE052206 and HV 5715 require 
dust shells of inner radius $\simali$17 and $\simali$52 times the stellar 
radius, respectively, with dust temperatures there of 900\,K and 430\,K, 
respectively, and with optical depths at 10$\mum$ through the shells of 0.095 
and 0.012, respectively.  The models compute the dust mass-loss rates for the 
two stars to be 2.0$\times$10$^{-9} \msunyr$ and 2.3$\times$10$^{-9} 
\msunyr$, respectively.  When a dust-to-gas mass ratio of 0.002 is assumed for 
SSTSAGE052206 and HV 5715, the dust mass-loss rates imply total mass-loss 
rates of 1.0$\times$10$^{-6} \msunyr$ and 1.2$\times$10$^{-6} \msunyr$, 
respectively.  These properties of the dust shells and stars, as inferred from 
our models of the two stars, are found to be consistent with properties observed 
or assumed by detailed studies of other O-rich AGB stars in the LMC and 
elsewhere.

\end{abstract}


\section{Introduction}

Asymptotic Giant Branch (AGB) stars are low- to intermediate-mass 
($\ltapp$ 8 $\msun$) stars that have reached the last stages of their 
lives as stars.  In this phase of an AGB star's life, the star expels its 
own circumstellar material, forming dust grains as the material 
moves away from the star, dragging the gas with it through 
momentum coupling.  The photospheric abundance of carbon 
relative to oxygen determines whether the ejected dust will be of 
oxygen-rich or carbon-rich composition \citep{hoef09}.  This ejected 
dust is subsequently added to the interstellar medium (ISM) 
surrounding the star.  At least some 
dust grains from AGB stars in our Galaxy survive their residence 
in the ISM and are incorporated into planet-forming disks around 
young stars, as must have happened for our Solar System 
\citep{gail09,nitt09}.

It is desirable to determine the relative contributions to the mass 
budget from different sources of dust in a galaxy.  Studies of 
dusty stars in our own Galaxy are difficult, as shorter-wavelength 
observations, necessary to constrain emission from the star and 
hot dust, are hampered by the high extinction along lines-of-sight 
through the disk of the Milky Way \citep[e.g., see][]{schul03}.  
Further, the often unknown extent of extinction by intervening dust 
hampers precise distance determinations, which, in turn, affects 
determining AGB star luminosities.  \citet{blom06} studied the mass 
loss of AGB stars in the Galactic Bulge; however, their surveys did 
not include lines of sight through the Galactic midplane, which can 
have very high extinction.  The {\it Surveying the Agents of a Galaxy's 
Evolution} (SAGE) {\it Spitzer} Space Telescope \citep{wer04} Legacy 
project was designed to study the life cycle of baryonic matter in the 
Large Magellanic Cloud \citep{meix06}.  Among other advantages 
of surveying the Large Magellanic Cloud (LMC) include a low 
average reddening of E(B-V)$\simali$ 0.075 \citep{schlegel98}.  
Also, because of the inclination angle of the LMC, all stars in the LMC 
are at roughly the same distance from us \citep[see discussion 
by][]{meix06}, which eases determination of their luminosities.

Infrared Array Camera \citep[IRAC;][]{faz04} SAGE observations 
found millions of stars in the LMC surveys.  Of the $\simali$32\,000 
of these millions that were classified as evolved stars brighter than 
the tip of the red giant branch (RGB) in the LMC by \citet{blum06}, 
over half ($\simali$17\,500) of them were found to be oxygen-rich 
(O-rich) AGB stars.  Another $\simali$7000 were found to be 
carbon-rich (C-rich) AGB stars, and $\simali$1200 were found to 
be ``extreme'' AGBs of undetermined chemistry.  Carbon-rich and 
extreme AGB stars amongst the SAGE sample will be explored in 
detail elsewhere \citep[see][]{srinphd09}; here, we focus instead on 
O-rich AGB stars.  The O-rich AGB stars were found to divide into 
two populations \citep{blum06}, a bright one and a faint one 
\citep[see also][hereafter Paper I]{srin09}.

To determine the relative contribution to the mass budget of the 
LMC from O-rich AGB stars, we desire to construct detailed 
radiative transfer (RT) models of each of these stars found in the 
SAGE survey.  However, these models require adequate dust 
optical properties in order to compute reliable dust mass-loss 
rates to be used in obtaining total mass-loss rates.  Numerous 
detailed RT studies have been conducted on O-rich AGB stars 
with an eye towards determining the optical properties of the 
silicate dust produced by such stars.  \citet{vokw88} used the RT 
code DUSTCD \citep{leu75,leu76a,leu76b,spleu83} to model 
AGB stars, constructing their own dust opacity function in the 
process, for which they found the more massive AGB stars 
typically had 10$\mum$ silicate features in absorption.  
\citet{schtie89} also used code by \citet{leu75,leu76a,leu76b} to 
model AGB stars, noting the classical problem of requiring more 
near-infrared wavelength (1 $<$ $\lambda$/$\mum$ $<$ 
8) continuum absorption than observed in typical terrestrial 
silicates to model the AGB stars successfully and proposing 
``color centers'' in astronomical silicates as a possible solution 
(among others).  \citet{simp91} constructed RT shell models to 
try to determine the dust emissivities for different groups of stars 
observed by the low-resolution spectrometer on {\it IRAS} to have 
silicate emission.  \citet{suh99} used CSDUST3 \citep{eg88} in 
modeling O-rich AGBs to construct optical constants sets for two 
different silicates - warm and cool, which were later used by 
\citet{suh04} to model the mass loss of pulsating AGB stars with 
low and high mass-loss rates.  \citet{kemp02} revisited the 
problem of the lack of near-infrared continuum opacity in AGB 
stars by modeling the OH/IR star OH 127.8+0.0 using the 
radiative transfer code MODUST \citep{bouw00,kemp01,bouwphd01}, 
concluding 
the near-infrared continuum arises from metallic iron.  Recently, 
\citet{heho05} have used the 1-dimensional RT code DUSTY 
\citep{ivez99} to model O-rich AGB stars with optically thin 
dust envelopes, finding evidence for various oxides and silicates.

IRAC and Multiband Imaging Photometer for {\it Spitzer} 
\citep[MIPS;][]{riek04} SAGE observations of some of the brightest 
sources were followed up with {\it Spitzer} Infrared Spectrograph 
\citep[IRS;][]{houck04} spectra as part of the {\it SAGE-Spectroscopy} 
(SAGE-Spec; PI: F. Kemper) {\it Spitzer} Legacy program 
\citep{kemp10}.  These spectra, along with IRAC and 
MIPS photometry, plus shorter-wavelength visible and near-infrared 
photometry obtained from other sources, allow detailed SEDs to 
be constructed for a small number of stars.  

Here, we determine dust grain properties that will allow reasonable 
radiative transfer model fits to the observed SEDs and spectra of O-rich 
AGB populations \citep{blum06}.  This search for typical dust properties 
is preparation for future work to determine the mass-loss contribution of 
O-rich AGB stars in the LMC to its total mass budget.  Ultimately, we 
wish to obtain the mass-loss rate for each of thousands of O-rich AGB 
stars in the LMC (found in the SAGE surveys) by radiative transfer 
modeling of its SED, which includes dust emission beyond 3.6$\mum$ 
wavelength.  This paper focuses on the radiative transfer modeling of a 
representative star from each of the bright and faint O-rich AGB 
populations, in order to find reasonable dust grain properties to use for 
more extensive later modeling of O-rich AGB stars.  A similar study has 
been undertaken for a C-rich AGB star (Srinivasan {et al., in prep}, 
hereafter Paper III).

\section{Observations}

As discussed above, \citet{blum06} noted the presence of two 
distinct populations of O-rich AGB stars in the SAGE sample of the 
LMC, a faint population and a bright population.  Allowing for 
differences in the dust properties between the populations, we chose 
a representative star from each population.  For the best constraints 
on the dust emissivity at mid-infrared wavelengths, we chose to model 
from each of the bright and faint populations a star with redder [8.0]-[24] 
colors than over half of its respective population, as redder colors 
suggest more prominent dust feature emission.  However, we did not 
want the [8.0]-[24] color to be too red, in order to avoid modeling an 
anomalous source.  Also, we wanted to model stars with SAGE-Spec 
IRS spectra, to provide tight constraints on the silicate emission features 
of our representative stars.  From the bright O-rich AGB star population 
we chose HV 5715 (SAGE-Spec ID 82).  From the faint O-rich AGB star 
population, identified 
by an ``F'' in the [24], [8.0]-[24] color magnitude diagram of \citet{blum06}, 
we chose SSTISAGE1C J052206.92-715017.6 (hereafter, SSTSAGE052206; 
SAGE-Spec ID 96).  The $K_{s}$ magnitudes for HV 5715 and SSTSAGE052206 
are 9.064 and 11.203, both well above the tip of the RGB of $K_{\rm s}$ = 
11.98 \citep{cioni00}, confirming the status of these stars as AGBs.  
These two stars have amongst the highest signal-to-noise (S/N) in the 
10 and 20$\mum$ silicate emission features of the O-rich AGB stars 
observed in the SAGE-Spec program.

Each of the two O-rich AGB stars we chose to model had $<$10\%  
agreement between the fluxes at 5.8, 8.0, and 24 microns synthesized 
from its 5--37$\mum$ {\it Spitzer}-IRS spectrum and the corresponding 
observed IRAC or MIPS {\it Spitzer} photometry.  The IRS spectra 
of these two stars show clear silicate emission, testifying to their 
O-rich nature.  In the 2MASS {\it K$_{\rm s}$}, {\it J}-{\it K$_{\rm s}$} 
color-magnitude diagram (CMD) of \citet{nw00}, SSTSAGE052206 
occupies region F, identified by \citet{nw00} as O-rich AGB stars of 
intermediate age.  The other star, HV 5715, was located in region 
G of the 2MASS CMD plotted by \citet{nw00}, corresponding to 
young AGB stars that have such a high mass that hot-bottom 
burning \citep{boosa92} prevents them from becoming or staying 
C-rich.

\subsection{Photometry and Variability}

Using MACHO data, \citet{wood99} found five sequences when 
plotting luminosity index versus log(P), where P is the period.  They 
identify sequence ``C'' with Miras and show the variability of the 
stars in this sequence to be consistent with pulsation in the 
fundamental mode.  With a larger dataset, \citet{fras05} resolved the 
sequence \citet{wood99} identified as ``B'' into two sequences, 
which they name sequences 2 and 3.  Both \citet{fras05} and 
\citet{fras08} show sequences 1 \citep[sequence C of][]{wood99}, 2, 
and 3 to have lightly populated, nearly vertical extensions at the 
bright ends of the sequences (the lowest $K_{\rm s}$ magnitudes).

MACHO data \citep{fras08} for HV 5715\footnote{see the MACHO 
light curves for MACHO id 49.6132.10 at the coordinates of HV 5715 
of Right Ascension 79.54623 degrees, Declination -67.4467 degrees 
available under ``Lightcurve Search'' at 
http://wwwmacho.anu.edu.au/Data/MachoData.html} indicate 
multi-periodic variability, with primary and secondary periods of 
415.97 and 211.06 days, respectively, with corresponding 
peak-to-peak MACHO blue-band amplitudes of 0.8 and 0.66, 
respectively.  On plots of $K_{\rm s}$ versus log(primary period) by 
\citet{fras08}, the point corresponding to the primary period of HV 5715 
falls within the edge of the nearly vertical extension to sequence 1.  The 
point corresponding to the secondary period of HV 5715 lies on the 
side of sequence 2 closest to sequence 3.  The plots of peak-to-peak 
amplitude versus log(period) and primary-to-secondary period ratio 
versus log(primary period) shown by \citet{fras08} are also consistent 
with the primary period of HV 5715 belonging to sequence 1.

SSTSAGE052206 matches the coordinates and average {\it V} and {\it I} 
magnitudes for OGLE-LMC-LPV-46603, identified as an ``OGLE Small 
Amplitude Red Giant'' (OSARG) in the Long Period Variable (LPV) list 
of the OGLE-III Catalog of Variable 
Stars\footnote{see http://ogledb.astrouw.edu.pl/$\sim$ogle/CVS/} \citep{sos09}.  
The catalog entry for OGLE-LMC-LPV-46603 gives primary, secondary, 
and tertiary periods of 81.24, 399.8, and 69.08 days, respectively, and 
their corresponding amplitudes (half of the peak-to-peak amplitude) for 
the light curve at I-band are 0.061, 0.047, and 0.038 magnitudes, 
respectively.  According to Fig. 2 of \citet{fras08}, its 
primary period and its $K_{\rm s}$ magnitude place it in between sequences 
3 and 2, slightly closer to 3 than to 2, while its secondary period places it 
between sequences 1 and D.  Both Figs. 8 and 9 of \citet{fras08} show 
SSTSAGE052206 to be very close to the ``one-year artifact'', which perhaps 
suggests some caution in the secondary period they find for this star of 
399.8 days.  \citet{fras08} note that OSARGs are closely related to RGB 
and E-AGB stars.  SSTSAGE052206 is only somewhat brighter than the 
tip of the RGB (see our earlier discussion), which suggests it to be in the 
earlier stages of its AGB star phase.

2MASS \citep{skrut06} {\it J}{\it H}{\it K$_{\rm s}$} and {\it Spitzer} 
IRAC 3.6, 4.5, 5.8, and 8.0 $\mum$ data for HV 5715 and 
SSTSAGE052206 come from the SAGE Winter 2008 IRAC Catalog, 
and {\it Spitzer} MIPS 24 $\mum$ data come from the SAGE Winter '08 
MIPS 24 $\mum$ Catalog.  Both catalogs are available on the 
{\it Spitzer} Science Center (SSC) 
website\footnote{http://ssc.spitzer.caltech.edu/legacy/sagehistory.html}.  
See \citet{meix06} and \citet{blum06} for more details on the SAGE 
epoch-1 point-source catalog, from which we obtain the data for 
HV 5715 and SSTSAGE052206.  We applied the zero-point offsets 
recommended by \citet{coh03} to the 2MASS data.  At shorter 
wavelengths, we use {\it UBVI} data from the Magellanic Clouds 
Photometric Survey \citep{zar97} for these two objects.  We correct 
the {\it UBVIJHK$_{\rm s}$} photometry for foreground extinction 
\citep[for more details, see][]{cioni06}.  The IRAC and MIPS photometry 
is not corrected for extinction, as the correction is negligible at those 
wavelengths.

To this photometry, for HV 5715 we add from \citet{glass79} 
{\it B}, {\it V}, $R_{ck}$, and $I_{ck}$ \citep[the last two being 
in the Cape Kron system; see][]{cousins76}, and two epochs of 
{\it J}, {\it H}, {\it K}, and {\it L} photometry.  In addition, we add 
{\it BVR} photometry from \citet{rebeirot83} and {\it J} and {\it H} 
photometry from IRSF \citep{kato07}.  To SSTSAGE052206 we 
add IRSF photometry at {\it J}, {\it H}, and $K_{\rm s}$ bands.  We 
correct this additional 
photometry (except for the {\it L} band flux) for extinction by 
linear interpolation in log(A$_{\lambda}$) versus log($\lambda$) 
space, where A$_{\lambda}$ is the extinction at wavelength 
$\lambda$, of the extinction law used by \citet{cioni06}.  Due to 
the multi-periodic nature of HV 5715 and both the multi-periodic 
nature and weak variability of SSTSAGE052206, we do not attempt 
to construct a single-phase SED for either star from multi-epoch 
photometry.  Instead, for each star we plot the photometry from 
all epochs on the same SED.

\subsection{Spectroscopy}

Data from the S15.3 and S17.2 pipelines for the Short-Low (SL) and 
Long-Low (LL) modules, respectively, were obtained from the SSC 
for SSTSAGE052206 (AOR \# 22422528) and HV 5715 (AOR \# 
22419456).  After reducing the spectra using techniques described 
by \citet{kemp10}, SL spectra were scaled up to match LL spectra 
in flux near 14.3$\mum$.

\section{Models}

\subsection{2Dust Radiative Transfer Models}

There are many RT codes from which to choose.  Many of the 
RT codes mentioned in the Introduction assume spherical symmetry.  
This is not as much a concern for very optically thin dust shells.  Every 
grain in such shells receives starlight with almost no extinction, and 
the radiation it scatters or emits thermally toward the observer 
likely experiences even less extinction due to the typically decreasing 
extinction efficiency of carbonaceous and oxygen-rich dust toward 
longer wavelengths.  However, our interests include very dusty AGB 
stars, for which cases the geometry of the circumstellar dust does 
play a strong role in the heating and emergent spectral energy 
distributions (SEDs) of the stars.  In anticipation of our future modeling 
of very dusty AGB stars that may have non-spherical dust shell 
geometries, we use the RT code {\bf 2D}ust \citep{um03}, which allows 
non-spherical axisymmetric circumstellar dust shell geometries.

\subsubsection{Dust Shell Geometry}

For the current study we simply assume spherical symmetry of the dust 
around the star, noting both the weakness of the mid-infrared flux relative 
to the flux at 1$\mum$ and the lack of silicate absorption features in the 
IRS spectra for both stars (see Figs. 1 and 2) suggest optically thin dust 
shells.  We assume a 1/r$^{2}$ density fall for the dust in the shell, which 
is expected for constant mass-loss rate.  The inner radius ($R_{\rm min}$) 
and outer radius ($R_{\rm max}$) are parameters that define the size of 
the shell.  $R_{\rm min}$ is varied for a best fit.  \citet{heho05} 
found the outer radius of the dust shell, $R_{\rm max}$, in their AGB models 
to be not easily constrained, and suggested a lower limit for their sample of 
100 times the inner radius of the dust shell.  We set the outer radius of the 
dust shell for both models at a thousand times the inner shell radius 
\citep{vokw88}, which is important for catching all the contributions 
from the dust shell to the 24$\mum$ MIPS flux.

\subsubsection{Stellar Temperature}

To represent the stellar photosphere emission, we use PHOENIX models 
\citep{all00} for stars of one solar mass and subsolar metallicity 
(log({\it Z}/$Z_{\rm Sun}$) = -0.5) to match determinations of the metallicity 
of the LMC \citep[{\it Z}/$Z_{\rm Sun}$ $\simali$ 0.3--0.5; see][]{west97}.  We 
favor PHOENIX models, as they include millions of lines from water vapor 
and other molecules critical for modeling cool, late-type stars as we do here 
with AGB stars.  
To give the best fit to the visible and near-infrared photometry, we use 
photosphere models corresponding to stellar effective temperatures, 
T$_{\rm eff}$, of 3\,500\,K $\pm$ 100\,K and 3\,700\,K $\pm$ 100\,K for HV 
5715 and SSTSAGE052206, respectively.  The uncertainty in the effective 
temperature was estimated to be about 100\,K for both stars, as PHOENIX 
models are given in 100\,K increments, and the photosphere models with the 
next highest and lowest effective temperatures gave marginally acceptable fits 
to the optical and near-infrared photometry.  Those stellar photosphere models 
with effective temperatures 200\,K greater or lesser than the ones we used 
provided noticeably worse fits to the optical and near-infrared photometry.  
We note here that selection of the best stellar photosphere 
model will be much more difficult when constructing RT models of highly 
optically thick dust shells for later model grids (Sargent et al., {\it in prep}) 
than for the two stars modeled here, as the optical and near-infrared colors 
will be affected by the optically thick dust shells.

\subsubsection{Stellar Luminosity}

PHOENIX models were only available for stars of one solar mass, but (as we 
discuss in \S4.5) our stars are more massive.  We therefore chose to use for 
modeling each of our two stars the one solar mass PHOENIX model with the 
nearest value of log(g) to what we estimate for the star.  To determine the 
correct value of log(g) to use for each star, where g is the 
gravitational acceleration at the star's surface in CGS units, we obtained a first 
guess for stellar radius and mass by assuming the star's mass is one solar 
mass.  We then adjusted the assumed stellar radius to a value that resulted in 
a good fit of the photometry from $U$- through $K_{\rm s}$-band wavelengths.  
From the luminosity of the resultant stellar photosphere and T$_{\rm eff}$, we 
placed our two stars on isochrones and determined stellar masses of $\simali$7 
and $\simali$3$\msun$ for HV 5715 and SSTSAGE052206, respectively (see 
discussion in \S4.5).  With stellar mass and our first guess at stellar radius, we 
determined values of log(g) for HV 5715 and SSTSAGE052206 of -0.15 and 
+0.43, respectively, so we used the PHOENIX models with log(g) values 
logarithmically nearest these values: 0.0 and +0.5, respectively.

To fit the observed photometry, we scale the fluxes of the T$_{\rm eff}$ = 
3\,500\,K, log(g) = 0.0 PHOENIX model for HV 5715 by 9.74, and we scale the 
fluxes of the T$_{\rm eff}$ = 3\,700\,K, log(g) = 0.5 PHOENIX model for 
SSTSAGE052206 by 3.48.  The luminosities of the resultant photospheres we 
use in our modeling are $\simali$36\,000 $\lsun$ $\pm$ 4\,000 $\lsun$ and 
$\simali$5\,100 $\lsun$ $\pm$ 500 $\lsun$ for HV 5715 and SSTSAGE052206, 
respectively.  We estimate the relative uncertainties on the luminosity for each 
of HV 5715 and SSTSAGE052206 to be about 10\%.  Should the fluxes for 
either star be scaled by more than 10\% from their current values, the fits to 
the overall SEDs and spectra would grow noticeably worse.  To obtain the 
stellar radii, this means scaling the radii of the unscaled PHOENIX models 
for HV 5715 and SSTSAGE052206 by $\sqrt{9.74}$ and $\sqrt{3.48}$, 
respectively.  Because L = $\sigma$T$^{4}$4$\pi$R$^{2}$, the 10\% relative 
uncertainties on the 
luminosities imply $\simali$5\% uncertainties on the stellar radii, assuming 
we have the correct stellar effective temperatures.  Supporting the assumption 
of our assumed subsolar metallicity is the fact that the only plot in Fig. 1 of 
\citet{marig08} for which both HV 5715 and SSTSAGE052206 would have 
effective temperatures and luminosities corresponding to O-rich AGB stars is 
the plot corresponding to Z=0.008.  This is about 0.42 times solar, consistent 
with \citet{west97}.  Here we note that if we scaled the stellar photosphere 
model for HV 5715 to fit the lowest fluxes in each band for which there is 
photometry from multiple epochs, we would obtain a luminosity of 
$\simali$31\,000 $\lsun$, about 15\% lower than the value we use (which 
instead fits the highest fluxes in the bands that have multi-epoch photometry).  
The stars are assumed to be 50 kiloparsecs away \citep{fea99}.

\subsubsection{Expansion Velocity ($v_{\rm exp}$)}

The dust is assumed to be moving away from from the star for each model 
at a terminal velocity of 10 km/s \citep[see][]{wood92,mar04}.  When we 
changed this parameter, it had no visible effect upon the SED, so we cannot 
constrain this parameter from our data.  The dust mass-loss rate is affected 
by this parameter, as it is linearly proportional to the expansion velocity by 
design of the {\bf 2D}ust code.  For HV 5715, the expansion velocity of 
10 km/s we adopted is consistent with Fig. 4 of \citet{marig08}, which plots 
expansion velocities for variable AGB stars versus their periods.  There are 
few data points with primary periods as low as that of SSTSAGE052206 in 
Fig. 4 of \citet{marig08}, but the three points with primary periods around or 
below 200 days have expansion velocities between $\simali$9 and 
$\simali$17 km/s, which are consistent with the value of 10 km/s we 
assume in our modeling.  The \citet{vw93} relation in that plot does not 
extend below primary periods of 200 days.

\subsubsection{Dust Cross-Sections and Sizes}

The {\bf 2D}ust models were run in 
Harrington averaging \citep{har88} mode, which means the dust 
cross-sections used to represent the dust properties in radiative transfer 
were cross-sections computed from weighted averages, with the weights 
being proportional to the grain surface area.  We assumed isotropic 
scattering because we found it to give output SEDs almost indistinguishable 
from those computed assuming anisotropic scattering \citep[using a 
modified Henyey-Greenstein phase function; see][]{cs92}.  Mie theory 
\citep{bh83} is used to compute the absorption and scattering 
cross-sections and asymmetry factor, {\it g}, of the dust grains, assumed 
spherical in shape, around the AGB stars.  However, real astrophysical 
dust grains are likely to be nonspherical \citep{bh83}.  The cross-sections 
of spherical and nonspherical grains differ most in the resonances 
(features), as ensembles of nonspherical grains tend to give wider features, 
with the long-wavelength side of the features pushed to longer wavelengths 
\citep[see][]{fab01,min05}.  This potential difference could cause a 
discrepancy between observed and modeled spectra.The dust grains were 
assumed to follow a KMH-like ``Power-law with Exponential Decay 
(PED)'' \citep{kmh94} grain size distribution, in which the number of grains 
of a given size is proportional to $a^{\gamma}$e$^{-a/a_{\rm 0}}$, where 
$a$ is the grain radius, $a_{\rm 0}$ sets the exponential decrease in number 
of grains to large sizes, and $a_{\rm min}$ is the minimum grain size.  When 
$a$ is much smaller than $a_{\rm 0}$, the grain size distribution acts 
approximately as $a^{\gamma}$, so $\gamma$ is fixed at -3.5 
\citep[after][]{mrn77}, and $a_{\rm min}$ and $a_{\rm 0}$ were allowed to be 
free parameters.  We found $a_{\rm min}$ = 0.01$\mum$ and $a_{\rm 0}$ = 
0.1$\mum$ to provide good fits of models to observed data.

\subsubsection{Fitting Procedure}

To obtain the best-fit model for each of our two stars, we generally began 
by determining stellar properties (stellar effective temperature and 
luminosity) first, dust shell properties next, and dust grain properties last.  
All other properties were fixed, as described previously in the text.  
Sometimes we had to iterate and loop through stellar, dust shell, and 
dust grain properties again.   After computing hundreds of models through 
such iteration, we found an acceptable combination of stellar, dust shell, 
and dust grain properties for both SSTSAGE052206 and HV 5715.

We would begin by selecting an unscaled 
PHOENIX stellar photosphere model of effective temperature and log(g) 
value that would give fluxes as close as possible to those of our stars.  We 
then scaled the photosphere models in flux up to match the SED fluxes, 
changing to using a PHOENIX model of differing log(g) value as needed 
(described previously).  Then the dust shell optical depth at 10$\mum$ 
($\tau_{\rm 10}$), was 
increased to the approximate level to match the flux in the 10 and 
20$\mum$ features.  Later, the dust shell inner radius ($R_{\rm min}$) 
was varied to obtain both the correct relative fluxes in the 10 and 20$\mum$ 
features and the correct slope of the underlying near- and mid-infrared 
continuum ($\lambda$ $<$ 8$\mum$ and 13$\mum$ $<$ $\lambda$ $<$ 
15$\mum$).  Lastly, $a_{\rm min}$ and $a_{\rm 0}$ were varied to try to 
improve the fit.  The fits were judged successful when the 10 and 20$\mum$ 
peak fluxes were matched, the near- and mid-infrared continuum in the 
model was as close as possible to that in the data, and the broadband 
fluxes at optical and near-infrared wavelengths were matched as closely 
as possible.  The details of the model properties are 
given in Table 1.  Figures 1 and 2, respectively, show the observed data 
and best-fit models for SSTSAGE052206 and HV 5715.

We note here that the IRS spectrum of HV 5715 may be well-fit by our 
model over its entire wavelength range (5 -- 38$\mum$) only if we fit 
just the maximum flux of each band for which there is photometry from 
multiple epochs.  This suggests the IRS spectrum for HV 5715 was 
obtained near its maximum in phase.  As this object has multiple periods 
and, as a result, has quite a complex light curve (see \S 2.1), we do not 
attempt to correct either the IRS spectrum or any photometry for 
phase.  We note the spread in fluxes seems to be small for the 
IRAC bands, grows slightly larger for {\it I}, {\it J}, {\it H}, and {\it K} 
bands, and grows larger still for the {\it V} and {\it B}, bands.  We further 
note that \citet{vijh09}, in their study of variability at mid-infrared 
wavelengths, list neither HV 5715 nor SSTSAGE052206 as variable, 
consistent with their infrared variability being relatively small.  
This increased amplitude in variability with shorter wavelengths may 
be intrinsic to the source.  As \citet{rg02} summarize, \citet{cel78}, 
building on earlier studies \citep[e.g.,][]{pn33,smak64}, noted the 
increasing amplitude in 
variability of Mira pulsating variables to shorter wavelengths.  
However, based on the low number of measurements we have in 
each of the concerned bands, we draw no further conclusions 
regarding the wavelength dependence of the variability of our stars.  
The single observed {\it U}-band flux we do have for HV 5715 (and 
SSTSAGE052206, for that matter) is higher than the flux synthesized 
from our model.  The reason for these {\it U}-band discrepancies is 
unknown, but it does support fitting only the maximum fluxes in 
the other bands of HV 5715, as fitting the median or mean of 
those bands would decrease the model flux at {\it U} even further 
below the observed flux than it already is.  The variability of 
SSTSAGE052206 is much smaller, so its model agrees quite well 
with both its photometry from all epochs and its IRS spectrum.

The dust shell inner radius ($R_{\rm min}$) and especially the dust shell 
optical depth at 10$\mum$ ($\tau_{\rm 10}$) have the greatest effects on 
the fluxes and colors of the output SEDs.  The grain size parameters, 
$a_{\rm min}$ and $a_{\rm 0}$, were of great interest, as we seek 
acceptable grain properties through modeling of HV 5715 and SSTSAGE052206 
to use in extensive radiative transfer modeling of O-rich evolved stars in the 
SAGE sample in the future.  All four parameters -- $R_{\rm min}$, 
$\tau_{\rm 10}$, $a_{\rm min}$, and $a_{\rm 0}$ -- were left free, to be 
determined by our radiative transfer modeling.  To gauge the range a parameter 
could vary and the overall fit of model to data remain acceptable, the fluxes 
longward in wavelength of 3$\mum$ were allowed to deviate by one to 
three times the uncertainties, as estimated by eye, while keeping all other 
parameters at their best-fit value (Table 1).  Figure 1 demonstrates this 
estimation of uncertainty by eye for the $\tau_{\rm 10}$ parameter for 
SSTSAGE052206, giving as orange curves the models obtained with 
$\tau_{\rm 10}$ set to the extremes of its allowable range (see values in 
parentheses in Table 1 beside the best-fit value for this parameter).  
The uncertainties on the other free parameters for SSTSAGE052206 and all four 
free parameters for HV 5715 were determined in the same way.  We list in 
Table 1 the uncertainties in $R_{\rm min}$, $\tau_{\rm 10}$, $a$, and 
$a_{\rm 0}$ determined by eye in parentheses beside their best-fit values.  
We save more exact determinations of uncertainties of model parameters 
and a detailed investigation of degeneracy of pairs of free model 
parameters for our future study of the entire model grid.  For now, we 
suggest the reader see \citet{speck09} for a further discussion on radiative 
transfer modeling parameter degeneracy and sensitivity to certain parameters.

\subsubsection{Dust Grain Composition}

Models using many different sets of refractory indices were constructed, 
but few were found to provide an overall good fit to the SED.  Models 
using refractory indices of amorphous silicates made from 
the ``sol-gel'' method \citep{jag03} were computed, but these silicates 
were found to achieve insufficient temperatures to match the infrared 
fluxes in the SEDs.  This was also true of the refractory indices 
of amorphous pyroxene of ``cosmic'' composition \citep{jag94} 
and of the amorphous pyroxenes of \citet{dor95} for stoichiometries with 
Fe/(Mg+Fe) values less than $\simali$0.5.  For the refractory indices 
for amorphous pyroxenes from \citet{dor95} with Fe/(Mg+Fe) 
values of 0.5 and 0.6, the dust temperatures were more reasonable 
and the model spectra more closely matched the observed spectra 
longward in wavelength of 8$\mum$, but the near-infrared continuum 
of the model shortward of 8$\mum$ was very much below the observed 
continuum at those wavelengths in the observed data (both IRAC 
photometry and IRS spectrum) for both HV 5715 and SSTSAGE052206.  
Also, the 20$\mum$ feature was too strong.  Both sets of amorphous 
olivine refractory indices from \citet{dor95} gave reasonable 
grain temperatures and reasonable fits to the IRS spectra longward of 
8$\mum$ and provided closer matches to the observed near-infrared 
continuum.  However, the extinction at {\it I} and {\it J} bands was too large for the 
model of SSTSAGE052206, and the 20$\mum$ features were still too 
strong compared to the IRS spectra.  The refractory indices of 
\citet{suh99} for ``warm'' and ``cool'' silicates and those of \citet{oss92} for 
oxygen-rich silicates were found to give moderately acceptable fits to the 
shapes of the 10 and 20$\mum$ emission features in the IRS spectra, but 
the peak-to-continuum ratio for the 20$\mum$ features in the models was 
too large, compared to the IRS spectra.

We found the refractory indices of oxygen-deficient silicates 
by \citet{oss92} to give the best fits of models to data, giving a physically 
reasonable dust shell geometry when fitting the models to the SEDs and 
giving the best overall fits to the 10 and 20$\mum$ features and to the 
near-infrared continuum in the spectra.  In order to use these constants, 
we had to add wavelength coverage, as the shortest wavelength of the 
oxygen-deficient silicate refractory indices of \citet{oss92} was 0.4$\mum$.  
To these \citet{oss92} refractory indices, we added indices between 0.2 
and 0.4$\mum$ determined by interpolating between the n and k values 
for \citet{oss92} at 0.4$\mum$ and the n and k values for the ``astronomical 
silicate'' of \citet{dl84} at 0.1718$\mum$.  
Increased average grain size would flatten the 10 micron feature 
and make it wider, pushing its long-wavelength side to longer wavelengths 
\citep[see][]{min05}, which is needed neither for HV 5715 nor (especially) 
for SSTSAGE052206.  In addition, the fits would not be improved by 
incorporating real nonspherical astrophysical grains because an 
ensemble of nonspherical grains will push the long-wavelength sides of 
the features to longer wavelengths \citep{min05}, which is not needed.  
Overall, the fits to the 10 and 20$\mum$ features for HV 5715 and 
SSTSAGE052206 are not perfect, 
but we are not aiming to fit the detailed shapes of the dust 
emission features in the spectra.  We also note our best fits are likely not 
unique; however, our goal here is to find good dust grain properties to 
use in radiative transfer modeling of O-rich AGB stars, with reasonable 
assumptions for the stellar properties and other dust shell properties.  
Instead, we aim to obtain good overall fits to the SEDs and save 
determination of details like grain shape distribution for future studies.

\subsubsection{Synthetic Photometry}

In order to compare our models, which have high wavelength resolution, 
to broadband photometry, we synthesized broadband fluxes, which we 
include in Figures 1 and 2 as diamonds, from our models.  We 
describe here how we synthesized photometry for the bands for which 
photometry was readily available in the SAGE catalogs; for the additional 
photometry we plot for HV 5715, we did not synthesize photometry from 
our model to compare.  For each of the bands of the {\it UBVI} photometry, 
we obtained the quantum-efficiency-based response function by multiplying 
the quantum efficiency (QE) of the Direct CCD Camera\footnote{Available at 
http://www.lco.cl/lco/telescopes-information/irenee-du-pont/instruments/specs/du-pont-telescope-direct-ccd-camera-ccd.  
This was extrapolated to 0.3$\mum$.  Also, the QE was assumed to 
linearly drop to zero from its value at 0.84$\mum$ wavelength, the 
last wavelength provided on the QE graph, to 1.13$\mum$, the 
wavelength corresponding to a photon energy of 1.1{\it eV}, which is the 
energy of the band gap of silicon.} by the transmission function of the 
band\footnote{The transmission functions of the {\it B} and {\it V} bands were 
obtained from http://www.lco.cl/lco/telescopes-information/irenee-du-pont/instruments/website/direct-ccd-manuals/direct-ccd-manuals/3x3-filters-for-ccd-imaging using the 
Harris {\it B} and Harris {\it V} filter profiles, which are LC-3013 and LC-3009, 
respectively.  The 
filter transmission profiles for {\it U} and {\it I} bands were obtained from I. 
Thompson (priv. communication).}.  The band fluxes for the {\it UBVI} bands 
were then obtained by computing the isophotal flux \citep[see Equation 
5 of][]{tv05}.  For each of the 2MASS {\it JHK$_{\rm s}$} bands, photon-counting 
relative spectral response (RSR) functions were 
obtained\footnote{http://www.ipac.caltech.edu/2mass/releases/allsky/doc/sec6\_4a.html} 
and used to compute isophotal fluxes \citep[see discussion in appendix 
E.4 of][regarding QE-based versus photon-counting response functions 
and computing band fluxes]{bess98}.  For the IRAC and MIPS photometry 
points, the fluxes were obtained using the IDL routine {\it 
spitzer\_synthphot}\footnote{For software and instructions, see 
http://ssc.spitzer.caltech.edu/postbcd/cookbooks/synthetic\_photometry.html}.

\subsection{SSTSAGE052206 Gas model}

A small feature appears in the spectrum of SSTSAGE052206 at 6.6$\mum$.  
A simple isothermal slab model of water vapor emission with a 
temperature of 1\,000\,K, column density of 10$^{18}$ cm$^{-2}$, and 
microturbulent velocity of 3 km/s, using a line list from \citet{ps97} 
convolved to R 
$\sim$ 90 was obtained using a model from the {\it spectrafactory} 
website\footnote{http://www.spectrafactory.net} \citep[see discussion 
by][]{cami10}.  We note the 6.6$\mum$ feature is present for water 
vapor of temperatures greater than 500\,K in these models, so the 
1\,000\,K temperature is not well constrained.  The assumed emitting 
surface was a circle whose emitting area is 2\,200$\rsun$ in 
radius ($\simali$13$\rstar$).  This is only $\simali$4 AU inward of the inner 
radius of the dust shell for this star (Table 1), and it supports the 
large dust shell inner radius we find from modeling the dust emission.

A carbon dioxide emission model was similarly constructed to fit 
the emission feature at 14.9$\mum$.  In studies of Galactic O-rich 
AGBs, it was suggested by \citet{just98} that stars with enhanced 
mass-loss rates would show little in the way of CO$_{2}$ emission.  
\citet{sloan03} showed that CO$_{2}$ emission strength in O-rich 
AGB stars was correlated with the strength of the 13$\mum$ dust 
feature, which was found to be stronger in semiregular variable 
stars than Miras.  There is no 13$\mum$ dust feature in SSTSAGE052206, 
and there are none of the other CO$_{2}$ lines either, suggesting 
weak CO$_{2}$ emission in SSTSAGE052206 more like that of a Galactic 
Mira star.  We obtained a model spectrum from the 
{\it spectrafactory} website to construct an isothermal slab model for 10$^{17}$ 
cm$^{-2}$ of CO$_{2}$ at 500\,K convolved to the same spectral 
resolution as the water vapor spectrum, using a line list from 
\citet{roth05}.  The emission was assumed 
to come from a circular area of radius 6\,000$\rsun$ ($\simali$35$\rstar$).  
This is qualitatively consistent with the picture presented by 
\citet{cami02} that the CO$_{2}$ feature originates further from the 
star than the H$_{2}$O feature.  H$_{2}$O and CO$_{2}$ 
features were also seen in the O-rich Mira star R Cas by 
\citet{mm00}, who find their results consistent with a model 
that includes a pulsation shock.

Being variable and near the maximum in its light curve may explain 
the CO$_{2}$ and H$_{2}$O features in the 
spectrum of SSTSAGE052206 being in emission.  At maximum, the greater 
luminosity of the star may heat up the circumstellar gas further 
away from the star.  This would result in emission from gas in 
front of a negligible background surpassing absorption by the 
parts of the same gas cloud (at the same temperature) that happen 
to lie in the line-of-sight between star and observer, resulting 
in net emission features from the gas \citep[for more, 
see][]{cami02}.  Indeed, the H$_{2}$O line-forming region is larger 
at maximum phase \citep[][]{matsu02}.  H$_{2}$O emission from 
the extended atmosphere and circumstellar gas fill the H$_{2}$O 
absorption from the photosphere.  However, because the photospheric 
H$_{2}$O has a higher excitation temperature than the H$_{2}$O in 
the outer atmosphere or circumstellar shell, and because photospheric 
absorption is so strong, H$_{2}$O emission from the outer shells is 
insufficient to fill the photospheric absorption completely 
\citep[][]{tsuji97}.  Indeed, the 6$\mum$ H$_{2}$O band is usually 
observed in absorption \citep{tsuji01}.  
\citet{cami02} did show that when the 14.9$\mum$ CO$_{2}$ feature 
was in emission, other gas molecules' features would also tend to 
be more in emission, hypothesizing more extended gas envelopes 
in these cases.  Perhaps SSTSAGE052206 has a very extended envelope, 
resulting in the water vapor features being in emission.

As can be seen from the gas emission models in Fig. 3, only 
water vapor contributes with any significance to the near-infrared 
continuum.  Even so, it contributes at less than the 10\% level, so 
the dust emission is responsible for most of the continuum emission 
in excess of that from the stellar photosphere shortward of 8$\mum$.  
To lower the near-infrared continuum to account for this minute 
water vapor emission shortward of 8$\mum$ would require either 
slightly increasing $R_{\rm min}$ or slightly decreasing any of 
$\tau_{\rm 10}$, $a_{\rm min}$, or $a_{\rm 0}$.  The contribution to 
continuum emission from CO$_{2}$ is negligible, so it is of no 
concern regarding better fitting the near-infrared continuum.

\section{Discussion}

\subsection{Dust Composition}

Figures 1 and 2 show that the models for both SSTSAGE052206 and 
HV 5715 provide overall acceptable fits to the observed spectra 
and SEDs.  The peak fluxes of the 10 and 20$\mum$ 
features are fairly well matched.  This is important, according to 
\citet{just05}, for deriving the mass-loss rate.  We also note the 
oxygen-deficient silicates by \citet{oss92} we used 
were designed to have similar optical properties to ``astronomical 
silicates'' empirically constructed to fit previous observations of 
AGBs and other astrophysical sources.  The problem 
of needing sufficient near-infrared ($\lambda$ $<$ 8$\mum$) 
continuum opacity has been identified for other O-rich AGB stars, 
such as OH 127.8+0.0 \citep[][]{kemp02}, WX Psc \citep[][]{dec07}, 
and HV 996 and IRAS 05558-7000 \citep[][]{vl99}.  It is likely related 
to the problem of astronomical silicates needing to be ``dirty'' 
(absorptive) enough to heat sufficiently \citep{schtie89}, and has 
been solved elsewhere by increasing the imaginary part 
of the complex dielectric constant over near-infrared wavelengths 
to increase the continuum opacity in that range \citep[e.g.,][]{dl84}.  
Since the \citet{oss92} silicates were based on previous empirically 
constructed ``astronomical silicates'', the good fit of the underlying 
continuum shortward of 8$\mum$ and between 13-15$\mum$ is 
somewhat expected.

The 10 and 20$\mum$ features in both model and data match 
very well in shape for HV 5715 and reasonably well for SSTSAGE052206.  
However, in detail the model and observed 10 and 20$\mum$ 
features of SSTSAGE052206 disagree slightly.  For SSTSAGE052206, the 10$\mum$ 
feature in the model peaks a few tenths of a micron longward in 
wavelength of the feature in the data, while the 20$\mum$ feature 
in the model peaks a few tenths of a micron shortward in 
wavelength of the feature in the data.  We have mentioned earlier 
the limitations of using Mie theory, which assumed spherical 
dust grains, and that real astronomical grains are likely not 
spherical \citep{bh83,fab01,min05}.  However, the use of an 
ensemble of nonspherical shapes would tend to push both 10 and 
20$\mum$ features to longer wavelengths \citep[e.g., see][]{fab01}.  
This would improve the fit to the 20$\mum$ feature of SSTSAGE052206, but 
it would worsen the fit to its 10$\mum$ feature.  Instead, we look to 
dust composition to explain these slight discrepancies in shape.  
Silicates with more silica-rich compositions, like pyroxenes, have 
10$\mum$ features centered at slightly shorter wavelengths than 
silica-poor silicates like olivines \citep{oss92}.  More silica-rich 
compositions also tend to have 20$\mum$ features shifted to 
longer wavelengths than silica-poor compositions 
\citep{dor95,jag03}.  This may suggest the silicates around 
SSTSAGE052206 may be slightly more silica-rich than those whose spectra 
were the basis of the ``astronomical silicates'' on which the 
\citet{oss92} silicates were based.  By extension, this suggests 
the SSTSAGE052206 silicates are more silica-rich than the HV 5715 
silicates.  We note, though, that we are modeling only two sources.  
We will explore this issue further in future papers.

Subtracting the stellar photosphere used 
in our models from the observed spectrum, HV 5715 has a 
classification of SE6/SE7, while SSTSAGE052206 has a classification of 
SE8 in the classification system used by \citet{sloan03}.  Both stars 
have what \citet{sloan03} call classic silicate emission.  More 
in-depth studies of the dust composition 
{\it via} the detailed spectral emission feature shapes await a 
future study.  However, the dust properties used here provide 
overall satisfactory fits to the SEDs and spectra of SSTSAGE052206 and 
HV 5715 and represent a baseline for a future study of the 
mass-loss rates of O-rich AGB stars in the LMC (Sargent {\it et al.}, in 
prep) by construction of large model grids covering a range of 
model parameters.

\subsection{Dust Temperature and Inner Radius}

SSTSAGE052206 and HV 5715 have dust temperatures at the innermost 
radii, $R_{\rm min}$, of their dust shells of 900\,K and 430\,K, 
respectively.  Based on the uncertainties of $R_{\rm min}$ for 
each of the two stars, the allowable range of temperatures at 
dust shell inner radius for SSTSAGE052206 is 1200\,K -- 700\,K, and the 
allowable range for HV 5715 is 650\,K -- 310\,K.  The dust 
temperatures at the dust shell inner radius for SSTSAGE052206 and HV 
5715 are comparable to the dust temperatures at innermost dust 
shell radius of 1\,000\,K \citep{bdjn87}, $\simali$900\,K 
\citep{schtie89}, and 400-700\,K \citep{simp91,suh04,heho05} in 
their radiative transfer models of O-rich AGB stars.  The dust shell 
inner radius of 17 $\rstar$ for SSTSAGE052206 is close to but just above 
the allowable range of dust shell inner radii for O-rich AGB stars 
(2.5-14$\rstar$) according to \citet{hoef07}.  Observations of O-rich 
AGB stars' dust shell inner radii indicate smaller radii of 3-6 $\rstar$ 
\citep{bes91,dan95}.  On the other hand, modeling by \citet{suh04} 
suggests low mass-loss rate O-rich AGBs (LMOAs) of similar 
luminosities to that of SSTSAGE052206 have much larger dust shell inner 
radii of 27 -- 41 $\rstar$.  As discussed at the end of \S3.2, if we 
were to include the water vapor emission in our model of SSTSAGE052206, 
we would need to lower the 5--8$\mum$ flux in the model slightly.  
Keeping $\tau_{\rm 10}$, $a_{\rm min}$, and $a_{\rm 0}$ 
approximately the same as their current values, this would mean 
$R_{\rm min}$ would need to increase slightly.

The dust shell inner radius for HV 5715 is 
52 $\rstar$, much larger than that for SSTSAGE052206.  \Citet{vl05} 
found the O-rich AGB stars and red supergiants with higher stellar 
effective temperatures and greater luminosities (or greater masses, 
comparing their Figs. 12 and 14) had larger dust-free 
inner cavities.  The stellar effective temperatures of HV 5715 and 
SSTSAGE052206 assumed for our modeling of 3\,500\,K and 3\,700\,K, 
respectively, are close, while their luminosities of $\simali$36\,000 
$\lsun$ and 5\,100 $\lsun$, respectively, suggest the relatively 
larger dust shell 
inner radius HV 5715 to be consistent with the \citet{vl05} result.  
Interferometric observations also tend to support more massive or 
luminous stars having relatively larger dust shell inner radii.  
\citet{ohnaka08} found the 40$\msun$ LMC red supergiant WOH 
G64 to have a dust shell inner radius of 15$\rstar$ and a dust 
temperature there of 880\,K.  Using 11$\mum$ interferometry, 
\citet{bes91} found the hottest dust around the Milky Way red 
supergiant $\alpha$ Ori to be $\simali$300\,K, and \citet{dan95} 
found $\alpha$ Ori and the supergiants $\alpha$ Sco and $\alpha$ 
Her to have dust shell inner radii near $\simali$38$\rstar$, also 
using 11$\mum$ interferometry.  The temperature of the dust at the 
inner radius of the dust shell for HV 5715 of 430\,K is lower than the 
expected condensation temperature of $\simali$900\,K according to 
\citet{schtie89}, but \citet{wada03} found amorphous silicates 
condensing at temperatures as low as $\simali$650\,K in laboratory 
experiments.  We note that it may be possible to raise the 
dust grain temperatures at the inner radius of the shell by decreasing 
the thickness of the dust shell \citep[][]{speck09}.  However, we further 
note from our earlier discussions that our model of HV 5715 is likely 
more characteristic of it during one of its maxima.  During minima in its 
light curve, HV 5715 is probably at least 15\% fainter (see \S3.1), so if 
the dust shell has not changed its dimensions, the temperature at the 
dust shell inner radius would be lower, due to the lower incoming flux 
from the star.

\subsection{Dust Mass-Loss Rates}

\citet{just04} found when constructing models of the SED of W Hya 
that the inferred mass-loss rate was not very sensitive to whether 
the dust properties were those of the amorphous pyroxene used 
in this study or the empirically-constructed ``astronomical silicates'' 
used by \citet{juti92} or \citet{ladr93}.  This study has found the output 
model SED fluxes to be fairly insensitive to $a_{min}$ and $a_{0}$, 
for the choice of \citet{oss92} complex indices of refraction used 
here.  The values for these parameters used in this study, 0.01 and 
0.1$\mum$, respectively, are chosen to be similar to ranges used 
in other modeling studies of AGB stars.  \citet{juti92} used grains 
in radius between 0.005 and 0.25$\mum$ to model OH/IR stars 
\citep[this range was also used by][to model the O-rich AGB star 
W Hya]{just04}.  \citet{kemp01} and \citet{kemp02} used a range of 
0.1-1$\mum$ to model O-rich AGB stars.

The dust mass-loss rates obtained from the {\bf 2D}ust models for 
SSTSAGE052206 and HV 5715 are, respectively, 2.0$\times$10$^{-9} {\rm 
\msunyr}$ and 2.3$\times$10$^{-9} {\rm \msunyr}$.  From the 
allowable ranges of optical depth, $R_{\rm min}$, and grain sizes, the 
dust mass-loss rate for SSTSAGE052206 could range from 1.1 -- 
3.3$\times$10$^{-9} {\rm \msunyr}$, and the same for HV 5715 could 
range from 1.1 -- 4.1$\times$10$^{-9} {\rm \msunyr}$.  If the shell 
expansion velocity for either of these two O-rich AGBs is different, the 
mass-loss rate would vary in a linearly dependent fashion on the 
expansion velocity.  Compared to dust mass-loss rates given in the 
literature for other O-rich AGB stars, the rates for SSTSAGE052206 and HV 5715 
are reasonable.  The dust mass-loss rates computed by \citet{juti92} 
for OH/IR stars with $\tau_{\rm 9.7}$ most similar to $\tau_{\rm 10}$ of 
$\simali$0.1 and $\simali$0.01 for the models of SSTSAGE052206 and HV 5715, 
respectively, in this study are quite similar.  According to \citet{juti92}, 
R Hor, R Cas, IRC+10523, and GX Mon have $\tau_{\rm 9.7}$ of 0.03, 
0.08, 0.13, and 0.13, respectively, and dust mass-loss rates of 0.62, 1.9, 
4.0, and 7.2 times 10$^{-9} {\rm \msunyr}$, though they assumed a 
constant luminosity for all stars of 10\,000\,$\lsun$.  \citet{schtie89} find 
a dust mass-loss rate for R Cas of 1.2$\times$10$^{-9} {\rm \msunyr}$ 
with $\tau_{\rm 10}$ of 0.10 and a luminosity of 29\,000$\lsun$.  For Z 
Cyg and o Ceti, \citet{suh04} find dust mass-loss rates 
between 0.76--1.6$\times$10$^{-9} {\rm \msunyr}$ for $\tau_{\rm 10}$ 
between 0.01--0.04 and luminosities between 4\,000 and 10\,000$\lsun$.  
Finally, using a computed 8$\mum$ excess emission of $\simali$7mJy 
for SSTSAGE052206 and $\simali$4mJy for HV 5715, Fig. 17 of Paper I 
suggests the dust mass-loss rates for SSTSAGE052206 and HV 5715 found in 
this study are a factor of $\simali$3 greater than expected from \citet{vl99} 
and a factor of $\simali$10 greater than expected from the empirical 
relation of 8$\mum$ excess emission and mass-loss rate given by 
Paper I.

\subsection{Inferred Total Mass-Loss Rates}

To translate dust mass-loss rates to total mass-loss rates, a dust-to-gas 
mass ratio must be assumed, but the values of such ratios can be fairly 
uncertain \citep[e.g.,][]{dec07}.  Various values quoted for oxygen-rich 
mass-losing stars include 0.01 \citep[for AGB stars;][]{kemp03}, 0.004 
\citep[for WX Psc, an AGB star;][]{dec07}, 0.003 \citep[for Miras;][]{juti92}, 
and 0.002 \citep[for VY CMa, a supergiant;][]{dec06}, among others.  
Here we assume a value of 0.002 to be consistent with Paper I.
\Citet{vl06} shows how the dust-to-gas mass ratio is linearly proportional 
to metallicity, Z.  If we assume the average dust-to-gas mass ratio of 
nearby O-rich AGB stars is 0.0063 \citep{knapp85,heho05}, and if we 
assume these nearby O-rich AGB stars have solar metallicity, then we 
would scale the dust-to-gas ratio by 0.4 (assumed metallicity of LMC; see 
section 3.1) to get $\simali$0.0025.  This is not far from our assumed 
dust-to-gas mass ratio.  For a given mass percentage of Al$_{2}$O$_{3}$, 
\citet{heho05} obtain a spread in gas-to-dust ratios that usually varies by a 
factor of 10 (see their Fig. 4), noting \citet{marengo97} found similarly 
large spreads in their gas-to-dust ratios.  We adopt this factor of 10 in the 
uncertainty of the dust-to-gas ratio we use.  Placing our value of 0.002 
logarithmically in the center of this range, we estimate our dust-to-gas ratio 
could be between 0.00063-0.0063.

The gas-to-dust ratio assumed here gives total mass-loss rates of 
1.2$\times$10$^{-6} {\rm \msunyr}$ and 1.0$\times$10$^{-6} {\rm \msunyr}$ 
for HV 5715 and SSTSAGE052206, respectively.  However, due to the large 
uncertainty we assign to the dust-to-gas ratio we assume, these values 
could range between 0.2--6.5$\times$10$^{-6} {\rm \msunyr}$ for HV 5715 
and 0.2--5.2$\times$10$^{-6} {\rm \msunyr}$ for SSTSAGE052206.  These total 
mass-loss rates are very close to each other, which is puzzling, given that 
the luminosities of the two stars differ by a factor of $\simali$7.  However, we 
note the uncertainties on the total mass-loss rates could also be consistent 
with the total mass-loss rates being a factor of $\simali$30 different between 
the two stars.  This does not take into account possible extra factors arising 
from the dust expansion velocities for the two stars differing in actuality from 
their assumed value of 10 km/s.  We do note, however, that even if the values 
of total mass-loss rate we find are correct, it could just be a coincidence 
arising from modeling only two sources.

The total mass-loss 
rates we find for HV 5715 and SSTSAGE052206 are comparable to the gas mass-loss 
rates (much larger than dust mass-loss rates, so approximately equal to the total 
mass-loss rates) determined from OH and CO observations of the four OH/IR 
stars - R Hor, R Cas, IRC+10523, and GX Mon - in the \citet{juti92} sample with 
the lowest optical depths (see their Table 2).  \citet{schtie89} find the total 
mass-loss rate for R Cas to be lower than but marginally consistent with the 
total mass-loss rates we find for HV 5715 and SSTSAGE052206.  Over half of the stars 
whose SEDs were modeled by \citet{heho05} have total mass-loss rates within 
the error bars of the rates for the two stars we model.  The total mass-loss rates 
we determine for our two stars are between the rates inferred for the high and 
low mass-loss phases of WX Psc, as determined by \citet{dec07}.  The mass-loss 
relation found by \citet{vl05} for O-rich AGB stars and red supergiants predicts 
total mass-loss rates of 8.6$\times$10$^{-6} {\rm \msunyr}$ for HV 5715 and 
7.8$\times$10$^{-7} {\rm \msunyr}$ for SSTSAGE052206, giving rates larger and smaller, 
respectively, than our {\bf 2D}ust modeling gives.

\subsection{Mass-Loss Rate and Evolutionary Status}

\subsubsection{HV 5715}

From the assumed temperature and luminosity we used in modeling 
the two O-rich AGB stars studied here, rough conclusions may be 
drawn about the two stars' natures.  According to the Hertzsprung-Russell 
(HR) diagram plotted in Fig. 12 of \citet{vl05}, the main-sequence 
progenitor of HV 5715 should have had a 
stellar mass of $\simali$7$\msun$, and, as such, it should be 
experiencing Hot Bottom Burning (HBB).  From mass-loss formalism for 
AGB evolution developed by \citet{vokw88}, the primary period of HV 
5715 of 415.97 days (see \S2.1) and its assumed luminosity from 
modeling suggest a stellar mass of just above 7$\msun$, consistent with 
the estimate based on the HR diagram of \citet{vl05}.  Further consistent 
with the high mass of HV 5715 is its location in Fig. 3 of \citet{groen94}, 
which suggests it to be more massive than the 5$\msun$ star for which 
tracks are plotted on the mass-luminosity plot.  Assuming an absolute 
bolometric magnitude, $\mbol$, for the Sun of 4.74 \citep{cox00} and 
assuming the luminosities used in our modeling, HV 5715 has $\mbol$ 
= -6.65, and SSTSAGE052206 has $\mbol$ = -4.53.  The closest data points to 
where HV 5715 would be plotted in Fig. 27 of \citet{groen09} are six 
stars for which HBB is inferred from Li detection, further supportive of HV 
5715 experiencing HBB.  Even further in support of the higher mass of HV 
5715 is its location in the $\mbol$ versus period plot (Fig. 8) of 
\citet{wood92}, which shows it to be located below the supergiants and 
above the AGB stars with no OH maser detections clustering around the 
4$\msun$ track.

We consider HV 5715 to be an AGB star and not a 
red supergiant (RSG).  \citet{wbf83} note the classical luminosity limit for 
an AGB star is at $M_{\rm bol}$ = -7.1; however, as \citet{sloan08} summarize, 
\citet{wood92} note a few AGB stars can occasionally move over that limit.  
\Citet{vl99} propose a dividing line between AGB and RSG of $M_{\rm bol}$ 
= -7.5, while \citet{groen09} use $M_{\rm bol}$ = -8.0 as a lower limit for 
the luminosity of stars they consider RSGs.  Expressing the distinction 
between AGB and RSG in luminosity, \citet{vl05} divide their sample of M 
stars by stating they consider the stars to be AGB if the luminosity is less 
than 10$^{4.9}$ (79\,433) $\lsun$, though they also note the classical AGB 
limit of $\simali$10$^{4.73}$ (53\,703) $\lsun$.  \citet{buch06} state the 
theoretical AGB luminosity limit to be 60\,000$\lsun$ but suggest AGBs 
can have luminosities slightly higher than this.  As we have already noted, 
HV 5715 would fall in region G of the $K_{\rm s}$ versus {\it J}-$K_{\rm s}$ 
color-magnitude diagram of \citet{nw00}, given its 2MASS {\it J}-$K_{\rm s}$ color of 
1.3, which they note would make it a massive (5--8$\msun$) AGB star.  This 
is consistent with the main-sequence progenitor mass we estimate for it of 
$\simali$7$\msun$.  \Citet{vl99} and others note that stars with initial masses 
greater than or equal to 8$\msun$ are typically considered RSGs.  This is 
consistent with Figure 9 of \citet{groen09}, which shows the 8$\msun$ 
evolutionary track to form an upper limit on the luminosities of the AGB stars 
in their sample, though there are a few AGB stars above this track and a few 
RSGs below the track.  HV 5715 would be marginally considered a RSG 
according to its {\it J}-$K_{\rm s}$ color and epoch 1 and 2 [3.6]-[4.5] colors (-0.054 
and -0.215, respectively), according to the {\it J}-$K_{\rm s}$ versus [3.6]-[4.5] 
color-color diagram in Figure 10 of \citet{buch09}.  However, in the color-color 
diagrams of Figure 9 of that paper, the [5.8]-[8.0] epoch 1 and 2 colors of 
0.157 and 0.187, respectively, and [8.0]-[24] epoch 1 and 2 colors of 1.556 
and 1.500 place HV 5715 outside of the region in each of the two diagrams 
labeled RSG, though it is on the side of the RSG region opposite the side 
adjacent to the region labeled ``O AGB''.  Caution is suggested, however, as 
the \citet{buch09} sample is only 250 sources.  The luminosity of 
36\,000$\lsun$, bolometric magnitude of -6.65, estimated mass of 7$\msun$, 
its colors, and its $K_{\rm s}$ magnitude suggest HV 5715 to be an AGB star, 
though these properties are also not far from being consistent with those of 
RSGs.

HV 5715 has a total mass-loss 
rate at least ten times higher than those of O-rich AGBs and RSGs of 
comparable [3.6]-[8.0] color (epoch 1, 0.397; epoch 2, 0.257), though its error 
bars make it marginally consistent with them.  This contrasts with HV 5715 
having a total mass-loss rate at least ten times {\it lower} than the rates for 
O-rich AGB stars of both similar luminosity in Fig. 8 of \citet{vl05} and similar 
$\mbol$ in Fig. 9 of \citet{vl99}, though, again, its error bars let it be marginally 
consistent with the rest of the O-rich AGB population.  We note here that the 
spread in mass-loss rates at a given luminosity \citep[e.g., as shown in Fig. 8 
of][]{vl05} is in part due to intrinsic differences between the stars at that luminosity 
and not only due to errors in determining the mass-loss rates.  As Fig. 13 of 
\citet{vl05} shows, mass-loss in evolved stars increases with both luminosity and 
stellar effective temperature.  In Paper I it is discussed how differences in 
mass-loss rate for stars at the same luminosity can be intrinsic.  To summarize, HV 
5715 has a total mass-loss rate that is high for its near-infrared color and low 
for its luminosity and bolometric magnitude, though the error bars on its total 
mass-loss rate make it marginally consistent with rates of O-rich AGBs and 
RSGs of similar near-infrared colors, luminosities, and bolometric magnitudes.

\subsubsection{SSTSAGE052206}

The HR diagram plotted in Fig. 12 of \citet{vl05} shows SSTSAGE052206 
to be more consistent with a $\simali$3$\msun$ star, so it is likely not 
experiencing HBB.  At such a mass, it may eventually become C-rich, 
which would be consistent with our earlier discussion suggesting it to 
be currently early in its AGB phase.  This lower mass for SSTSAGE052206 
is also consistent with the \citet{groen94} period-luminosity relation (their 
Fig. 3), which shows the $\mbol$ of SSTSAGE052206 of -4.53 to be lower 
than all the tracks of the 5 $\msun$ star and in the middle of the tracks of 
the 1.25 $\msun$ star.  \citet{vokw88} do not plot lines for primary periods 
as low as that of SSTSAGE052206 of 81.24 days in their Fig. 9, but 
approximate extrapolation of their curves down to the primary period of 
SSTSAGE052206 suggests a stellar mass below 4.5$\msun$.  The 
primary period for this star is below the range plotted in Fig. 27 of 
\citet{groen09}, but it looks in that plot to be consistent with O-rich AGB 
stars with no Lithium detected, which implies SSTSAGE052206 is not 
experiencing HBB.

The mass-loss rate for SSTSAGE052206 and its bolometric magnitude place 
it in the middle of the region populated by C-rich AGB stars according to Fig. 9 
of \citet{vl99}; however, the error bars on its mass-loss rate also make it consistent 
with a couple of M-type AGB stars in the same figure.  This is also true of Fig. 8 
of \citet{vl05}, which, in addition, shows SSTSAGE052206 to be also consistent, 
in terms of total mass-loss rate and luminosity, with two of the three MS or S-type 
stars plotted in that figure.  This similarity to S- or C-type AGB stars in these plots 
may also be consistent with the location for SSTSAGE052206 in the Z=0.008 
(approximately the LMC metallicity assumed here; see \S3.1) plot of Fig. 1 of 
\citet{marig08}.  This location in the \citet{marig08} HR diagram suggests that in 
the future, SSTSAGE052206 may become a carbon-rich AGB star.  In Fig. 24 of 
\citet{groen09}, the total mass-loss rate of SSTSAGE052206 is quite consistent 
with the total mass-loss rates of O-rich AGB stars and red supergiants (RSGs) of 
similar [3.6]-[8.0] color (its epoch 1 color is 1.376 and its epoch 2 color is 1.301).  
Figure 21 of \citet{groen09} shows SSTSAGE052206 to be most consistent in its 
{\it I}-band pulsation amplitude and total mass-loss rate with a number of red 
supergiants and a couple of C-rich AGB stars in the Small Magellanic Cloud.  To 
recap, SSTSAGE052206 has a total mass-loss rate consistent with other O-rich 
AGBs of similar near-infrared color and consistent with stars of mixed chemistry 
(M-, MS-, S-, and C-types) of similar luminosities, bolometric magnitudes, and 
{\it I}-band pulsation amplitudes.

\section{Conclusions}

In this study, we constructed detailed radiative transfer models using {\bf 2D}ust 
of the SEDs of two O-rich AGB stars in order to find useful dust properties to use 
in later modeling of the entire O-rich AGB population found by the SAGE surveys 
of the LMC.  A similar study has determined useful dust properties to use in later 
modeling of the C-rich population in the SAGE LMC surveys (Paper III).  We 
chose to model one star each from the bright and faint 
populations of O-rich AGB stars identified by \citet{blum06}.  We required each 
star to have an IRS spectrum from the SAGE-Spec program in order to model in 
detail the silicate emission features.  We further required each star to have a red 
[8.0]-[24] color relative to the rest of the O-rich AGB population, so that the dust 
emission was prominent at mid-infrared wavelengths; however, we did not want 
the [8.0]-[24] colors to be too red, lest we choose an outlying, unusual star.  From 
the bright O-rich AGB population we chose HV 5715, and from the faint O-rich 
AGB population we chose SSTSAGE052206.

We have fitted the photometry and spectroscopy of these two stars with {\bf 2D}ust 
SED models using the same dust properties.  These properties include the use of 
complex indices of refraction of oxygen-deficient silicates \citep{oss92}, a 
``KMH''-like grain size distribution (for more, see \S3.1.5 ``Dust Cross-Sections and 
Sizes'') with a power-law exponent, $\gamma$, of -3.5, 
$a_{\rm min}$ of 0.01$\mum$, and characteristic size for exponential tail-off to 
large sizes of $a_{0}$ = 0.1$\mum$.  These dust grain properties represent a 
baseline set to use for constructing future grids of models of O-rich AGB stars to 
compare to the observed SAGE data.  We note that the dust properties we used 
provide an excellent fit to the SED and spectrum of HV 5715, the bright O-rich AGB 
star, after taking into account its photometric variability.  The fit to the SED and 
IRS spectrum of SSTSAGE052206, the faint O-rich AGB, is good overall.  We do 
note the 10 and 20$\mum$ features in the model peak longward and shortward, 
respectively, of the features in the data, suggesting the silicates of SSTSAGE052206 
to be more silica-rich than both those of the oxygen-deficient \citet{oss92} silicates 
we use and those of HV 5715.  This possible difference in dust composition will be 
sought in future studies of O-rich AGB stars in the LMC.

Simple models of water vapor and carbon dioxide gas emission for SSTSAGE052206 
suggest negligible contribution to the near-infrared continuum flux from the 
carbon dioxide and at most 10\% of the flux between 5--8$\mum$ to come from 
water vapor.  This suggests our modeling of the near-infrared continuum flux 
arising completely from dust emission is not a bad approximation.  Were we to 
include gas emission in our dust models, we would need either to increase 
$R_{\rm min}$ slightly or to decrease any of $\tau_{\rm 10}$, $a_{\rm min}$, or 
$a_{\rm 0}$ slightly.  We also find the large region from which the water vapor 
arises to be consistent with the large dust shell inner radius we find for SSTSAGE052206.

We derive $\mdot$ = 2.3$\times$10$^{-6} 
{\rm \msunyr}$ for HV 5715 and $\mdot$ = 2.0$\times$10$^{-6} {\rm \msunyr}$ for 
SSTSAGE052206, although the error bars on each of these of 0.2--6.5$\times$10$^{-6} 
{\rm \msunyr}$ and 0.2--5.2$\times$10$^{-6} {\rm \msunyr}$, respectively, are very 
large.  We note these uncertainties arise from the uncertainties in dust mass-loss 
rates and the dust-to-gas mass ratios assumed.  The former originate from 
uncertainties estimated by eye for the free parameters $R_{\rm min}$, 
$\tau_{\rm 10}$, $a_{\rm min}$, and $a_{\rm 0}$, which were likely conservatively 
estimated - the formal uncertainties on these parameters may be smaller.  The 
latter uncertainty, that in the dust-to-gas mass ratio, contributing to the uncertainty 
in total mass-loss rate would be much smaller if we had detailed measurements 
of the gas component of these stars' circumstellar shells.  We have not factored 
in the uncertainty in the dust expansion velocity.  We believe the value we have 
assumed for dust expansion velocity is close to the real value for HV 5715.  We 
have no information on the dust expansion velocity for SSTSAGE052206.

The total mass-loss rates we find for HV 5715 and SSTSAGE052206 may have 
very large uncertainties, but we note that the purpose of this study was not to 
obtain precise total mass-loss rates; instead, we seek good dust properties to use 
in later modeling of the entire O-rich AGB population of the SAGE surveys.  By 
fitting a the SED and spectrum of a star from each of the bright and faint populations 
of O-rich AGB stars \citep{blum06}, we intend to account for possible differences 
between the average dust properties of the two populations.  We find good 
fits of our models to the data for these two stars.  We also note that a major goal 
of SAGE is to find the relative rates of dust injection from the different sources of 
dust in the LMC, which includes red supergiants, C-rich AGB stars and ``extreme'' 
AGB stars in addition to the bright and faint populations of O-rich AGB stars.  Such 
does not require knowledge of total mass-loss rates, only dust mass-loss rates, 
which means modeling of the dust emission from such evolved stars will be 
sufficient to achieve that goal.  We find that the total and dust mass-loss rates, 
dust shell inner radii and temperatures at those radii, and other modeling 
parameters are consistent with those assumed and inferred by other studies of 
O-rich AGB stars with similar properties to each of HV 5715 and SSTSAGE052206.  
This gives confidence that the dust properties we find for these two stars will be 
useful in modeling the rest of the O-rich AGB stars in the SAGE sample.

\acknowledgements This work is based on observations made with 
the {\it Spitzer Space Telescope}, which is operated by 
the Jet Propulsion Laboratory, California Institute of Technology 
under NASA contract 1407.  This publication makes use of the Jena-St. 
Petersburg Database of Optical Constants \citep{hen99}.  The authors 
wish to thank the anonymous referee for comments that greatlly improved 
this manuscript.  The authors would also like to thank Kevin Volk, Sacha 
Hony, Albert Zijlstra, Jacco van Loon, and Martha Boyer for helpful 
comments and discussion.  We wish to thank Peter Hauschildt for his 
assistance with the PHOENIX stellar photosphere models.  The authors 
have made use of the SIMBAD astronomical database and would like to 
thank those responsible for its upkeep.  The authors also would like to 
thank Bernie Shiao at STScI for his hard work on the SAGE database and 
his kind assistance.

\clearpage

\begin{figure}[t] 
 \epsscale{0.8}
 \plotone{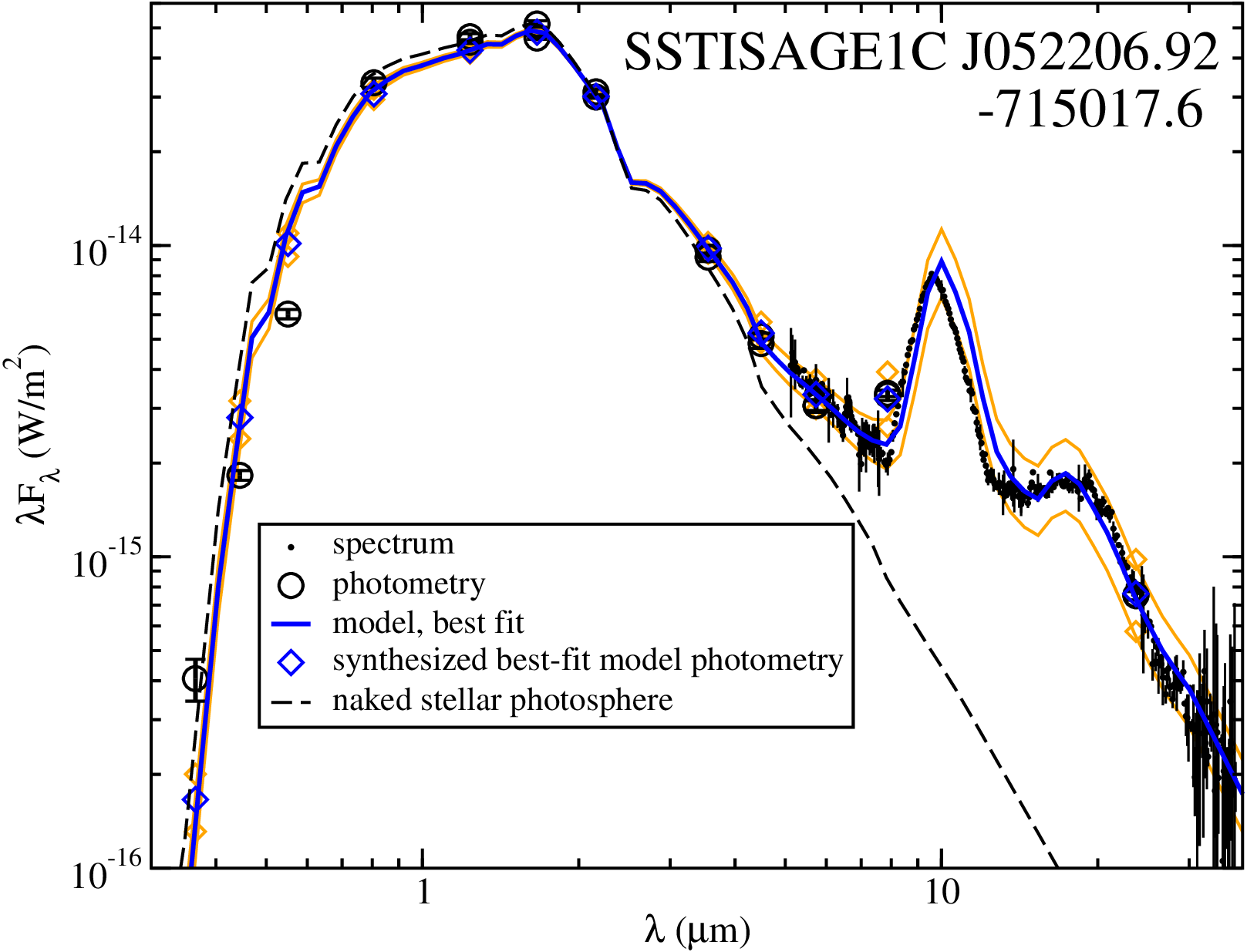}
 \caption[SSTSAGE052206]{{\bf 2D}ust model fit to the SED of SSTSAGE052206.  
 The small black dots with vertical lines through them are the 
 IRS spectrum data points with error bars, the large open 
 circles with errorbars are the observed photometry (see \S2.1 
 for sources of photometry), the model fit is the blue thick solid 
 line, the large open blue diamonds are the 
 photometry synthesized from the model, and the black dashed 
 line is the naked stellar photosphere model.  Components in 
 orange are the model and synthetic photometry for models 
 with optical depth at 10$\mum$, $\tau_{\rm 10}$, at the extremes 
 of its allowable range.  The lower orange curve and set of 
 points corresponds to $\tau_{\rm 10}$ = 0.07, and the upper 
 orange curve and set of points corresponds to $\tau_{\rm 10}$ = 0.125.  
 This demonstrates how the uncertainty for $\tau_{\rm 10}$ was 
 determined.}
\end{figure}

\clearpage

\begin{figure}[t] 
 \epsscale{0.8}
 \plotone{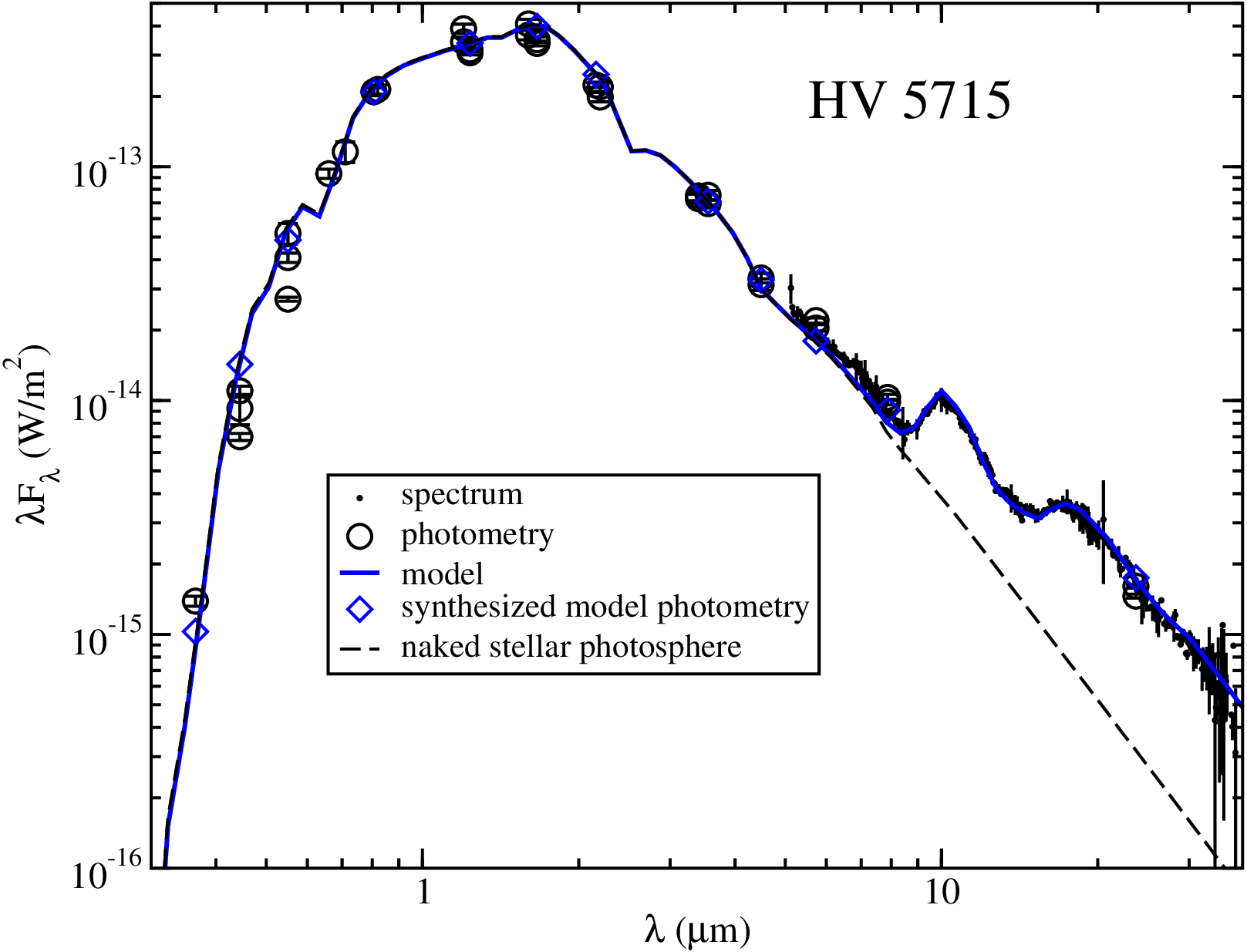}
 \caption[hv5715]{{\bf 2D}ust model fit to the SED of HV 5715.  
 Same meaning of symbols as for Fig. 1, without the orange 
 curves and points.}
\end{figure}

\clearpage

\begin{figure}[t] 
 \epsscale{1.0}
 \plotone{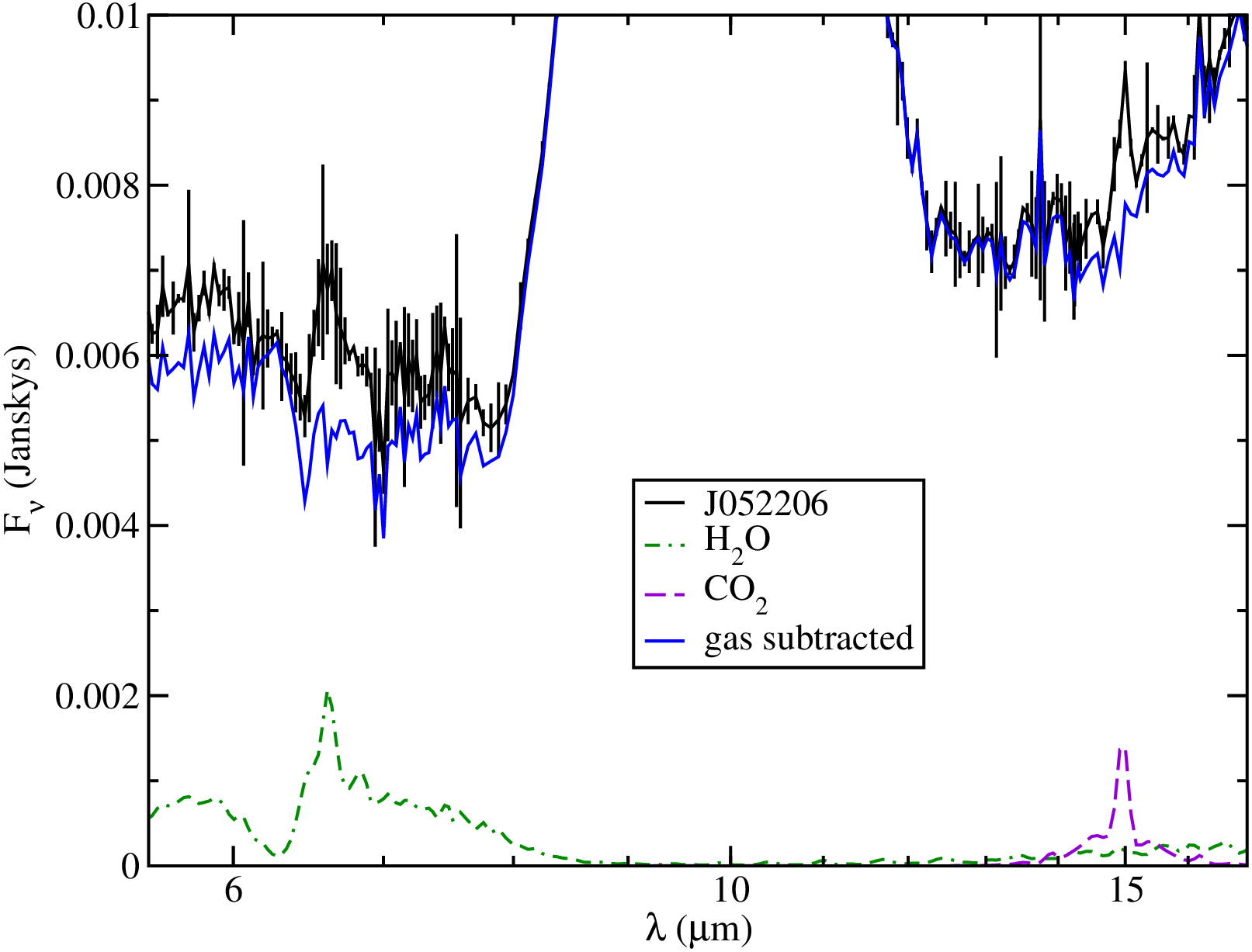}
 \caption[gas]{The spectrum of SSTSAGE052206 is compared to models 
of CO$_{2}$ and H$_{2}$O gas emission.  The H$_{2}$O model is 
the green dot-dash-dash line, the CO$_{2}$ model is the purple 
dashed line, the IRS spectrum of SSTSAGE052206 is the solid black 
line with errorbars, and SSTSAGE052206 spectrum minus both gas 
models is the solid blue line with no errorbars.}
\end{figure}

\clearpage

\begin{table}[h,t]
{
\caption[2Dust Parameters]{{\bf 2D}ust Model Parameters and Results\label{table1}}
\begin{tabular}{lcccccc}
\hline \hline
          
          &
          & SSTISAGE1C\\
          
          & HV 
          & J052206.92\\
          
          & 5715
          & -715017.6\\
\hline
{\bf Star} & & \\
$T_{eff}$ (K) & 3500 $\pm$ 100 & 3700 $\pm$ 100\\
Log(g) & 0.0 & 0.5\\
Log($Z/Z_{sun}$)* & -0.5 & -0.5\\
$R_{star} (\rsun)$ & 520 $\pm$ 30 & 170 $\pm$ 10\\
$L_{star} (\lsun)$ & 36\,000 $\pm$ 4\,000 & 5\,100 $\pm$ 500\\
{\bf Dust Grains} & &\\
$\rho_{dust}$* (g/cm$^{3}$) & 3.3 & 3.3\\
$\gamma$* & -3.5 & -3.5\\
$a_{min} (\mum)$ & 0.01 (0.0003, 0.08) & 0.01 (0.0003, 0.09)\\
$a_{0} (\mum)$ & 0.1 (0.02, 0.3) & 0.1 (0.02, 0.5) \\
{\bf Assumed Values} & &\\
$R_{max}$/R$_{min}$* & 1000 & 1000\\
$v_{exp}$* (km/s) & 10 & 10\\
{\bf Dust Shell} & &\\
$\tau_{10}$ & 0.012 (0.009, 0.015) & 0.095 (0.07, 0.13)\\
$R_{min} (\rstar)$ & 52 (25, 93) & 17 (9, 28)\\
$T_{d,inner}$ (K) & 430 (310, 650) & 900 (700, 1200)\\
$\mdot_{dust} (10^{-9} \msunyr)$ & 2.3 (1.1 -- 4.1) & 2.0 (1.1 -- 3.1)\\
$\mdot_{total} (10^{-6} \msunyr)$ & 1.2 (0.2 -- 6.5) & 1.0 (0.2 -- 5.2)\\
\hline
\end{tabular}
\tablecomments{\footnotesize The photosphere model flux for HV 
5715 was obtained by scaling the flux from the original photosphere 
model (which had log(g) of 0.0) by 9.74 and 
corresponding R$_{star}$ by $\sqrt{9.74}$; for SSTSAGE052206, the original 
photosphere model (with log(g) of 0.5) flux was scaled by 3.48 and 
R$_{star}$ by $\sqrt{3.48}$.  A KMH grain size 
distribution n(a) $\propto$ a$^{\gamma}$e$^{-a/a_0}$ \citep{kmh94} 
was used for both models.  An asterisk (*) indicates a parameter was 
fixed, not determined from model fitting.  Values in parentheses beside 
the best-fit values for $\tau_{10}$, $R_{min}$, $a_{min}$, and $a_{0}$ 
are the allowable ranges of uncertainty of these parameters, as described 
in subsections of \S3.1.  Note also the gas-to-dust mass ratio assumed, 500, 
for computing the total mass-loss rate from the dust mass-loss rate is quite 
uncertain, as discussed in the text.}
}
\end{table}

\clearpage

\begin{table}[h,t]
{
\caption[Gas Model Parameters]{SSTSAGE052206 Gas Model Parameters\label{table2}}
\begin{tabular}{lcc}
\hline \hline
          
          & H$_{2}$O
          & CO$_{2}$\\
\hline
$T_{\rm gas}$ (K) & 1000 & 500\\
$N_{\rm gas}$ (cm$^{-2}$) & 10$^{18}$ & 10$^{17}$\\
$v_{\rm turb}$ (km/s) & 3 & 3\\
$R_{\rm slab} (\rstar)$ & 13 & 35\\
\hline
\end{tabular}
\tablecomments{\footnotesize $T_{\rm gas}$ is the temperature of 
the isothermal gas slab, $N_{\rm gas}$ is the gas column 
density into the slab, $v_{\rm turb}$ is the microturbulent velocity of the gas, and 
$R_{\rm slab}$ is the radius (expressed in stellar radii; for the stellar radii, see 
see Table 1) of the emitting area, assumed circular, of the slab.  The gas 
models are convolved to a spectral resolution of 90.}
}
\end{table}

\end{document}